\documentclass{article}
\usepackage{pdflscape}
\usepackage{makeidx}
\usepackage{comment}
\usepackage{authblk}
\usepackage{hyperref}
\usepackage{graphicx}
\usepackage{caption}
\usepackage{subcaption}
\usepackage{longtable}
\usepackage{makecell}

\usepackage{color}
\usepackage[usenames,dvipsnames,svgnames,table]{xcolor}
\usepackage{booktabs}
\usepackage{threeparttable}
\usepackage{appendix}
\usepackage{textcomp}
\usepackage{gensymb}
\usepackage[round]{natbib}
\usepackage{csquotes}
\usepackage{mathtools}
\usepackage{amssymb}
\usepackage{amsmath,amsthm}
\usepackage{xfrac}
\usepackage{enumitem}
\usepackage{rotating}
\usepackage{verbatim}
\hypersetup{
backref=true, 
bookmarks=true, 
unicode=false, 
pdftoolbar=true, 
pdfmenubar=true, 
pdffitwindow=false, 
pdfstartview={FitH}, 
pdftitle={}, 
pdfauthor={}, 
pdfsubject={}, 
pdfkeywords={}{}, 
pdfnewwindow=true, 
colorlinks=true, 
linkcolor=BlueViolet, 
citecolor=NavyBlue, 
urlcolor=blue,
}
\title{Confidence Intervals for Recursive Journal Impact Factors}
\author[a]{Johannes K\"{o}nig}
\author[b]{David I. Stern}
\author[c d e f g h]{Richard S.J. Tol}
\affil[a]{International Center for Higher Education Research (INCHER) and Department of Economics, Economic Policy, Innovation and Entrepreneurship, University of Kassel, Germany}
\affil[b]{Arndt-Corden Department of Economics, Crawford School of Public Policy, The Australian National University, 132 Lennox Crossing, Acton, ACT 2601, Australia, david.stern@anu.edu.au}
\affil[c]{Department of Economics, University of Sussex, BN1 9SL Falmer, United Kingdom, r.tol@sussex.ac.uk}
\affil[d]{Institute for Environmental Studies, Vrije Universiteit Amsterdam, The Netherlands}
\affil[e]{Department of Spatial Economics, Vrije Universiteit Amsterdam, The Netherlands}
\affil[f]{Tinbergen Institute, Amsterdam, The Netherlands}
\affil[g]{CESifo, Munich, Germany}
\affil[h]{Payne Institute for Public Policy, Colorado School of Mines, Golden, CO, USA}
\begin{document}

\maketitle

\begin{abstract}
We compute confidence intervals for recursive impact factors, that take into account that some citations are more prestigious than others, as well as for the associated ranks of journals, applying the methods to the population of economics journals. The \textit{Quarterly Journal of Economics} is clearly the journal with greatest impact, the confidence interval for its rank only includes one. Based on the simple bootstrap, the remainder of the ``Top-5'' journals are in the top 6 together with the \textit{Journal of Finance}, while the \citet{xie2009}, and \citet{mogstad2022} methods generally broaden estimated confidence intervals, particularly for mid-ranking journals. All methods agree that most apparent differences in journal quality are, in fact, mostly insignificant.
\end{abstract}





\textbf{Acknowledgements}: We thank Carl Bergstrom for sending us information on the Eigenfactor metrics and Daniel Wilhelm for help with the ranks bootstrapping. Johannes K\"{o}nig gratefully acknowledges support from the German Federal Ministry of Education and Research (BMBF) under grant number 16PU17007A and from Deutsche Forschungsgemeinschaft (DFG) (FOR 5234).

\newpage
\section{Introduction}
\label{sc:intro}
Academics are all too human in their love of pecking orders. Ranking journals is a popular pastime and a key step towards ranking researchers and departments. Unfortunately, as impact factors are stochastic measures of underlying journal quality \citep{stern2013}, journal rankings create an unwarranted sense of precision. Therefore, it is important to understand when distinctions between journals may be reliably made and when they cannot. \citet{stern2013} computed confidence intervals and other statistics for \emph{simple} journal impact factors (IFs) for all economics journals included in the \emph{Web of Science}. In this paper, we propose a method to compute the standard errors of \citet{pinski1976} \emph{recursive} impact factors \footnote{This method is also known as the "invariant method" for recursive IFs. \citet{palacios2004} show that the \citet{pinski1976} method is axiomatically the best recursive impact factor method.} and their ranks, which we apply it to the population of economics journals.

Recursive impact factors reflect not only how many times a publication has been cited but also where it was cited, with greater weight placed on citations in outlets that are themselves highly cited. Where the simple impact factor assumes that all citations are created equal, the recursive impact factor acknowledges that some citations are more prestigious than others. Recursive IFs better reflect journal prestige, while simple IFs reflect popularity \citep{bollen2009}. Because economists tend to evaluate the research of other economists based more on where their work was published rather than how many citations it has received, the most popular journal rankings in economics use recursive or iterative impact factors \citep{liebowitz1984,laband1994,kalaitzidakis2003,kodrzycki2006,palacios2004,zimmermann2013,ham2021},  though there is evidence that total citation numbers best predict salaries \citep{hamermesh2018} and the distribution of economists across institutions \citep{sterntol2021}. 

\citet[][188-189]{stern2013} notes that ``simple IFs may not be the most appropriate measure of journal quality and recursive indicators are more popular in economics journal ranking studies. However, there is no simple way to construct uncertainty measures for these iterative indicators. It seems likely that the level of uncertainty concerning rankings revealed here would also attend rankings produced using more sophisticated indicators.'' While simple impact factors are means, the recursive impact factors are an eigenvector. We, therefore, use bootstrapping to compute confidence intervals and standard errors for recursive journal IFs. We also compute confidence intervals for the journals' ranks. Inference on ranks is complicated too, as the distribution of a rank is distinctly non-normal and the distributions of ranks are not at all independent.\footnote{\citet{mogstad2022} compute confidence intervals for the ranks of economics journals using Stern's data. \citet{horrace2017} use the same data to assign journals to best and second-best "clubs" while controlling for multiple comparisons.} We use a simple bootstrap, as suggested by \citet{goldstein1996},\footnote{While \citet{goldstein1996} suggest using bootstrapping to derive confidence intervals for ranks of different populations, they do not actually use this approach in their paper. \citet{hall2009} simply state that bootstrapping ranks is popular and convenient. Nevertheless, we refer to this as the Goldstein method in many places.} and adapt the methods of \citet{xie2009}, and \citet{mogstad2022} to directly use our article level data. The results are similar to those for simple IFs reported by \citet{stern2013}: \textit{The Quarterly Journal of Economics} is clearly the top ranked journal.  Based on the simple bootstrap, the remainder of the ``Top-5'' journals are in the top 6 together with the \textit{Journal of Finance}, while the \citet{xie2009}, and \citet{mogstad2022} methods generally broaden estimated confidence intervals. All methods agree that most apparent differences in journal quality are, in fact, mostly insignificant.

\citet{stern2013} uses the standard error of the mean to compute the standard error for simple IFs, as the simple IF is a mean: The average number of citations in the year that we compute the impact factor for \textemdash the ``citation year'' \textemdash to each article published by the journal in question in the "publication window" we are assessing. Its standard error is the standard deviation of the number of citations received in the citation year divided by the square root of the number of articles published in the publication window. So, for example, to compute the 2019 5-year IF, we compute the average citations received in 2019 by articles published in a given journal in 2014-18. To compute the IF’s standard error, we use the same list of the number of times each of the 2014-18 articles was cited in 2019.

If instead we used bootstrapping to compute this standard error, we would analogously resample with replacement from the list of 2014-18 articles and each time compute the IF for that new sample. The standard deviation of this IF is the estimated standard error. This approach follows logically because this is the standard way to compute a bootstrapped standard error, we sample from the observed data used to construct the statistic in question. We do not construct a counterfactual where the numbers for the observations change. The latter would be the case, if instead we re-sampled the citing articles so that the number of citations each article received changed. 

Recursive IFs are based on a matrix of how many times each journal was cited by each other journal in the citation year \textemdash the ``journal cross-citation matrix''. Underlying this matrix is a matrix of how many times each article published in the publication window was cited by each journal in the citation year. For example, each row could be an article published in 2014-18 and each column list how many times each of those articles was cited by a particular journal in 2019.

We propose to bootstrap recursive IFs in a similar way to the simple IF:
\begin{enumerate}
    \item Reesample with replacement from the list of all articles published in the publication window but make sure to sample the same number of articles from each journal as there are in the observed data.
    \item Compute the matrix of how many times each journal cited each other journal, the cross-citation matrix, from this.
    \item Compute the recursive impact factors from this and associated matrices.
    \item Use the distribution of values from repeated runs to construct the standard errors, confidence intervals etc. for each journal.
\end{enumerate}

We also construct confidence intervals for journal ranks by computing the ranks of the journals in each run and by combining this bootstrapping approach with the \citet{xie2009} and \citet{mogstad2022} methods for computing the confidence intervals of ranks.

\citet{lyhagen2020} is the only previous research on this topic that we are aware of. They use a bootstrapping method to compute confidence intervals for the Pinski and Narin recursive IFs. Their method only uses the journal cross-citation matrix and so does not resample articles. They redistribute the references in one of the journals to the others randomly using the probabilities represented by the entries of the cross-citation matrix and then recalculate all the ranks. Using these changes in ranks, they estimate confidence intervals for the recursive IFs. Therefore, their method misses the variation of citations \emph{within} journals, which is likely greater than the variation of citation \emph{across} journals \citep{oswald2007, wall2009}. In our sample, in a regression of citation counts of articles on journal fixed effects, the fixed effects only explain 16\% of the variation in the number of times articles are cited. On the other hand, our analysis (in common with \citet{lyhagen2020}) misses citations to and from journals in other disciplines. \citet{aistleitner2019} show that the top 5 journals in economics tend to cite each other more than those in other social science disciplines do. Assuming this extends to economics journals in general, it is more reasonable to limit our analysis to only economics journals than it would be if we were examining, for example, sociology journals\textemdash but this choice does, of course, disproportionately affect journals positioned at the boundaries of economics.

The original purpose of computing journal citation metrics was to inform librarians about which journals they should subscribe to \citep{Garfield1972}. Given that most citations were to a small subset of all journals, subscriptions could focus on these. However, journal level citation metrics are often used to evaluate individual research papers and individual researchers. There has been much criticism of this use of journal level citation metrics \citep{waltman2020} culminating in the \href{https://sfdora.org/}{San Francisco Declaration on Research Assessment}. As the distribution of citations to the articles in any journal is usually very dispersed and skewed \citep{seglen1992}, the correlation between journal impact factors and the citations received by individual articles is necessarily low. \citet{oswald2007} finds that the best article in an issue of a good to medium-quality journal routinely goes on to have much more citation impact than a ‘poor’ article published in an issue of a more prestigious journal. However, this does not mean that there is no correlation between impact factors and paper citations \citep{lozano2012} and there is also a strong negative correlation between IFs and journals’ acceptance rates, which measure journal selectivity and, therefore, are a proxy for quality \citep{aarssen2008}. 

\citet{waltman2020} develop a model where researchers try to submit their paper to the most prestigious journal in terms of its impact factor, but citations are an imprecise measure of underlying paper quality, and the review process is also stochastic. Then depending on the relative variances of the review and citation processes, journal impact factors can be a better measure of the underlying quality of a paper than the actual number of citations a paper receives. More generally, journal citation metrics are useful for assessing recently published and forthcoming papers, which have not yet had time to receive many citations \citep{abramo2010,levitt2011, stern2014}. However, because of the high variance of citations to articles in individual journals a measure of the uncertainty of impact factors is essential.

The paper continues as follows. Section \ref{sc:methods} presents the data (\ref{sc:data}), reviews alternative approaches to computing recursive IFs (\ref{sc:rif}), and discusses how to carry out bootstrapping (\ref{sc:bootstrap}), compute the recursive impact factors in practice (\ref{sc:compute}), and methods for computing confidence intervals for ranks (\ref{sc:confidence}). Section \ref{sc:results} presents the results and Section \ref{sc:conclusion} concludes.

\section{Data and Methods}
\label{sc:methods}

\subsection{Data}
\label{sc:data}
Our initial sample includes all 323 journals in the economics category in the \textit{Journal Citation Reports} that have citable items in all of the years 2014, 2015, 2016, 2017, 2018, and 2019.\footnote{Unlike \citet{ham2021}, we do not attempt to identify which of these are really economics journals and which are not.} We obtain the number of citations in each of the 323 journals in 2019 to the citable items published in each journal in the five year period, 2014-2018 via our access to the Clarivate database. The Clarivate database drives the well-known \textit{Web of Science}, which includes papers published in a ``rigorously selected core of journals'' and their forward and backward citations.  We include articles and reviews only, as the sum of these corresponds to the number of citable items in the \textit{Journal Citation Reports}. This difference between citable items and total items is greatest for the journal \textit{Value in Health}, where the most common publication type is meeting abstracts.

We dropped the following journals that gave a low number of citations in 2019 to articles published in 2014-18 in the other journals. As a result, it was possible to select no articles from these journals which referenced other journals in the bootstrapping. In that case, the matrix $D$ in (\ref{eq:rif}) is non-invertible and our algorithm crashes. The following journals were excluded:
\begin{itemize}
 
	\item \textit{Australian Journal of Political Economy} (ISSN: 0156-5826), which gave only 11 citations in 2019 to articles published in 2014-18 in the other journals. 
	\item \textit{CEPAL Review} (ISSN: 0251-2920), which gave only 7 citations in 2019 to articles published in 2014-18 in the other journals.
    \item \textit{Revue d’Etudes Comparatives Est-Ouest} (ISSN: 0339-0599), which gave only 4 citations in 2019 to articles published in 2014-18 in the other journals.
	\item \textit{Journal of Competition Law and Economics} (ISSN: 1744-6414), which gave only 7 citations in 2019 to articles published in 2014-18 in the other journals.
\end{itemize}

There are 319 journals and 88,928 articles in total in the remaining dataset.

We sorted the data in Excel by the ISSN of the cited journal and then by cited article to make the subsequent computation easier. The data files that we use in our analysis consist of the list of the details of the articles, including the journal in which they were published, a file containing the number of articles published in each journal, and a matrix of the number of times each article was cited by each journal. We wrote programs in RATS and Matlab to carry out the computations. The output data from the bootstrapping consist of a vector for each journal of the value of the recursive IF in each of the 1000 iterations from which we compute the results in Table \ref{tab:results}. These vectors  are numbered in order of ISSN.

\subsection{Recursive Impact Factors}
\label{sc:rif}
\citet{palacios2004} show that the \citet{pinski1976} ``invariant'' method is axiomatically the best recursive impact factor method. It obtains the impact factors as the positive eigenvector (Perron–Frobenius theorem), $v$, of the following matrix:
\begin{equation}
\label{eq:rif}
  A^{-1} C D^{-1} A  
\end{equation}
where $C$ is the journal cross-citation matrix of entries $C_{ij}$, which denotes the citations received by journal $i$ from journal $j$. $D$ is a diagonal matrix with $C_j=\sum_i C_{ij}$, the total citations that journal $j$ gave to other journals (or the sum of its references), on its diagonal. $A$ is a diagonal matrix with the number of articles published in each journal, $A_i$, in the publication window on its diagonal. Each of these matrices has dimensions $J\times J$, where $J$ is the number of journals.

Underlying $C$ is a matrix $P$ $(K\times J)$ of the number of times each article published in the publication window was cited by each journal in the citation year. $K$ is the total number of articles published in all journals in the publication window including articles with no citations. Each row of the matrix is for an article published in the publication window and each column is for a citing journal. So $P_{k,j}$ is the number of times journal $j$ cited article $k$ in the citation year. Matrix $Q$ $(J\times K)$ has a row for each cited journal and a column for each article and an entry of one for $Q_{i,k}$ indicates that article $k$ was published in journal $i$. Then:
\begin{equation}
\label{eq:ccalc}
C = Q P
\end{equation}

Note that $A^{-1} C$ is the average number of citations that an article in journal $i$ got in journal $j$ as each of its entries is $\sfrac{C_{ij}}{A_{i}}$. The sum of each row of $A^{-1} C$ is, therefore, the simple IF of journal $i$, $\sfrac{C_i}{A_{i}}$, where $C_i=\sum_j C_{ij}$. The entries of $D A^{-1}$ are $\sfrac{C_j}{A_j}$, where $A_j$ is the number of articles in the citing journal, which can be called the reference intensity of journal $j$, the average number of references in each paper. Therefore, Equation (\ref{eq:rif}) normalizes $\sfrac{C_{ij}}{A_i}$\textemdash the impact of journal $i$ on journal $j$\textemdash by $\sfrac{C_j}{A_j}$ \textemdash the reference intensity of journal $j$. This normalization is important. Some journals insist on extensive literature reviews. Other journals impose strict limits on the number of words and references. To be included in a short reference list is more impressive than to be included in a long one. 

By contrast, \citet{liebowitz1984} compute the positive eigenvector of $A^{-1} C$ and so do not adjust for the reference intensity of the citing journals. \citet{kalaitzidakis2003}, \citet{laband1994}, and \citet{kalaitzidakis2011} use the \citet{liebowitz1984} method in their rankings of economics journals and departments. On the other hand, \citet{kodrzycki2006}, \citet{Bao2010}, \citet{lo2016}, \citet{lyhagen2020}, and \citet{ham2021} use the Pinski and Narin method.

Alternatives to the invariant index have been proposed in order to provide various appealing properties \citep{koczy2013,demange2014}. \citet{palacios2014} show that the handicap method \citep{demange2014} satisfies a different set of axioms than satisfied by the invariant method. It is unclear whether one is better than the other. The modified invariant method of \citet{koczy2013} replaces the matrix $A$ with $D$ in order to address the potential for journals to manipulate rankings by publishing fewer articles, as they argue that both the Liebowitz and Palmer and invariant method are biased against "article splitting", Note that the fewer articles a journal publishes the wider its confidence interval will be, \textit{ceteris paribus}.

The \textit{Scimago Journal Rank} \citep{gonzalez2010} uses an iterative approach. \citet[][p. 380]{gonzalez2010} argue that Pinski and Narin’s method ``presented problems that were essentially related to the topological structure of the citation network''. First, each journal is assigned the same initial prestige value $\sfrac{1}{J}$, where $J$ is the number of journals in the database. Then the iterative procedure begins. Each iteration assigns new prestige values to each journal in accordance with three criteria: (1) a minimum prestige value from simply being included in the database; (2) a publication prestige given by the journal’s share in the number of papers included in the database; and (3) a citation prestige given by the number and ``importance'' of the citations received from other journals. This size-dependent measure is then normalized to give a size-independent metric, the SJR indicator. In order to prevent excessive journal self-citation, the number of references that a journal may direct to itself is limited to a maximum 33\% of its total references. These rules are \textit{ad hoc} and lack the elegance and axiomatic justification of the method of \citet{pinski1976}.

The Eigenfactor metrics of \citet{bergstrom2010} start with the matrix $C$, setting the diagonal to zero to eliminate self-references and then normalizing the columns to sum to one. The article vector, $a$, is the shares of each journal in total articles. They replace any zero columns of the normalized $C$ by this. Call the resulting matrix $H$. Then they find the positive eigenvector, $\pi$, of
\begin{equation}
   E = \varrho H + (1-\varrho) a \textbf{1}' 
\end{equation}
where $\textbf{1}$ is the unit vector. They then compute $H\pi$ and normalize the resulting vector to sum to one. This is theorized to give the percentage of time a random-walking reader would spend at each journal. It is analogous to the ``Google matrix'' that Google uses to compute the PageRank scores of websites. The stochastic process can be interpreted as follows: for a fraction $\varrho$ of their time, the random walker follows citations and for a fraction $1-\varrho$ of their time the random walker ``teleports'' to a random article chosen at a frequency proportional to the number of articles published. The Article Influence score, reported in the Journal Citation Reports, divides this fraction by the number of articles a journal published in the publication window.

\citet{gleich2009} examines the sensitivity of PageRank to the parameter $\varrho$. Writing the PageRank vector as a function of $\varrho$ allows him to take a derivative with respect to $\varrho$ as a simple sensitivity measure. Gleich found that PageRank is very sensitive to the choice of $\varrho$ and yet there is no obvious way to choose  $\varrho$, \textit{a priori}. Of course, this source of uncertainty does not appear in the Pinski and Narin formulation.

\subsection{Bootstrapping}
\label{sc:bootstrap}

The bootstrapping procedure iterates the following steps $B=1000$ times:
\begin{enumerate}
    \item For journal $i=1$ to $i=J$, compute $A_i$ random integers between 1 and $A_i$, where $A_i$ is the number of papers published in journal $i$, using a uniform distribution and place in vector $r$.
    \item Select articles from the sample with replacement by forming the matrices $P_b$ and $Q_b$, analogous to the empirical $P$ and $Q$, and setting $P_{b,k,j} = P_{r_k,j}$ and $Q_{b,i,k} = Q_{i,r_k}$.
    \item Compute $C_b = Q_{b} P_{b}$. This takes the bulk of the computation time of each iteration. We also compute a new version of the normalization matrix, $D_b$. The article matrix, $A$, does not change using the cluster bootstrap.
    \item Compute $v_b$ as the positive eigenvector of ${A}^{-1} {C_b} {D_b}^{-1} A$.
    \item Save $v_b$.
    \end{enumerate}
We then use the distribution of the $v_b$'s to estimate the distribution of $v$ and its associated statistics.

We use a cluster bootstrap, as in \citet[][Section 5]{efron1986}, by separately resampling from the data for each journal. That is because each journal has its own distribution of citations, so that not only the mean number of citations \textemdash the simple impact factor\textemdash varies across journals but also the variance of citations and the higher moments vary. This also ensures that the number of articles in each journal in each draw is the same as in the empirical data. However, this makes the assumption that the data drawn from each journal is an independent random sample and that samples from each journal are independent of each other. In essence, individual draws in the bootstrap procedure ask what would happen to the recursive impact factor if papers with particular citation profiles had not been published and other papers with other citation profiles had been published twice or more. The assumption of independence between journals disregards that if a paper had not been published in, say, \textit{Econometrica}, it would probably have been published in the \textit{Journal of Econometrics}. However, as we only have anecdotal information on submission histories, we cannot estimate the dependence between journals and that correlation would not be invariant to paper quality.

Alternatively, we could resample with replacement from the whole set of articles in all journals, which assumes that all articles are drawn from a single distribution. However, this means that the number of articles in a journal is not constant across draws. We tested these two approaches on a pilot sample of the top five journals only and it made no discernible difference to the results.\footnote{A more radical alternative is as follows. In our proposed bootstrap, we randomly draw from the set of \textit{cited} articles, assessing what would have happened to the recursive impact factor had a slightly different set of papers been published. There is another source of variation \textemdash we could have drawn from the set of \textit{citing} articles instead. Then the data set would consist of the number of times each citing article cited each cited journal. This moves the uncertainty away from the journal for which we estimate the recursive impact factor to all journals. This is conceptually awkward if we see the recursive impact factor as an input into the decision of authors about where to submit their paper\textemdash surely, authors do not submit to a journal in order to boost the impact of other journals. Similarly, editors decide to publish a paper if they think it is suitable for their journal, not because of the papers it cites. Numerically, sampling from \textit{citing} articles would lead to variation in citation numbers for the \textit{cited} articles that is smaller than for sampling from \textit{cited} articles.} 

\subsection{Computing the recursive impact factors}
\label{sc:compute}
We use the power iteration method to find the positive eigenvector. To estimate the empirical recursive impact factors, we initialize the process using the vector of simple impact factors $v_0=A^{-1} C \textbf{1}$, where $\textbf{1}$ is a vector of all ones, which sums the rows of $C$. Each iteration $t$ then follows:
\begin{equation}
\label{eq:power}
    v_t = \frac{Vv_{t-1}}{|Vv_{t-1}|}
\end{equation}
where $V=A^{-1} C D^{-1} A$. Dividing by the Euclidean length of the vector ensures that $|v_t|=1$ and so the final vector of recursive IFs, $v$, has length one. We found that 20 iterations were sufficient for the vector of recursive impact factors to converge from the simple IFs. 

To estimate the recursive impact factors from the bootstrapped data, we initiate the process using the empirical recursive IFs and substituting $C_b$ and $D_b$ for $C$ and $D$ in (\ref{eq:power}). We found that 10 iterations were sufficient for convergence from the empirical recursive IFs to the recursive IFs for the bootstrapped sample. 

\subsection{Confidence intervals for ranks}
\label{sc:confidence}
In addition to computing confidence intervals, standard errors, and other quantiles for the recursive IFs, we compute confidence intervals for the ranks of each journal using three alternative methods (plus two variants). First, we use the rankings in each realization of the bootstrap to construct a 95\% confidence interval of the rank, as proposed by \citet{goldstein1996}. This is straightforward: For each realization of the bootstrap, we rank journals. Across realizations, we find the 2.5 and 97.5 percentiles of the ranks for each journal.

\citet{xie2009} show that computing confidence intervals for ranks by simply computing the ranks of each journal on each iteration and compiling the distribution of ranks for each journal across all iterations leads to misleading inference when values of the statistic of interest for different journals are tied or close to tied. Confidence intervals based on the simple bootstrap are too narrow, as rankings are overconfident. To see this, note that the rank $R_i$ of journal $i$ can be computed as
\begin{equation}
\label{eq:goldstein}
    R_i = 1 + \sum_{j \neq i} I_{v_j > v_i} + 0.5 \sum_{j \neq i}  I_{v_j = v_i}
\end{equation}
where $v_i$ is the estimate of the recursive impact factor of journal $i$ and $v_j$ of journal, and $I$ is the indicator function equal to one if the subscripted condition is true and zero otherwise. \citet{xie2009} propose to replace the indicator function in Equation (\ref{eq:goldstein}) by a symmetric, standardized cumulative distribution function, such as the standard Normal distribution, $F$
\begin{equation}
\label{eq:xie}
    R_i^X = 1 + \sum_{j \neq i} F \left ( {\frac{v_j - v_i}{\tau_{i,j}}} \right )
\end{equation}
where $\tau_{i,j}$ is the ``bandwidth''. If $v_i$ and $v_j$ are far apart, $F$ assigns 0 or 1, and if $v_i = v_j$ $F$ assigns 0.5, just like the indicator function does. However, Equation (\ref{eq:xie}) avoids spurious precision when $v_i$ and $v_j$ are close but not the same. ``Close'' is defined by the bandwidth $\tau_{i,j}$. If $\tau_{i,j} \rightarrow 0$, $R^X \rightarrow R$. Following \citet{xie2009}, we set $\tau_{i,j} = \gamma \hat{\sigma}_{i,j}^\beta$ where $\gamma$ is the interquartile range of the original data that we are ranking and $\beta=0.5$. However, \citet{xie2009} set $\tau_{i,j} = \tau_i$ independent of $j$. Departing from \citet{xie2009}, we instead use $\tau_{i,j} = \hat{\sigma}_{i-j}$, the bootstrap standard deviation of the \emph{difference} in the recursive impact factors of journals $i$ and $j$ \textemdash the measure of distance used by \citet{mogstad2022}\textemdash rather than just the standard deviation of $v_i$. 

\citet[][Theorem 2.1]{xie2009} show that the bootstrap confidence interval of the smoothed rank is a consistent estimator of the true confidence interval in the presence of ties. \citet[][Theorem 2.3]{xie2009} shows in the presence of near ties that the confidence interval based on bootstraps of Equation (\ref{eq:goldstein}) is inconsistent, essentially because the indicator function is discontinuous. They further show that a consistent estimator of the confidence interval of the rank, based on Equation (\ref{eq:goldstein}), widens the bootstrap confidence interval (of the smoothed ranks) by 
\begin{equation}
\label{eq:xiecorr}
    T_i =  \sum_{i \neq j} f \left ( {\frac{v_j - v_i}{\tau_{i,j}}} \right )  
\end{equation}
where $f$ is a kernel function, e.g. the standard Normal density. $0.5T_i$ is subtracted from the lower bound and added to the upper bound of the bootstrap confidence interval.\footnote{There is a problem with Equation (\ref{eq:xiecorr}): If $\tau \rightarrow 0$, $f \rightarrow \infty$. We therefore set $f \leq 1$, so that any tie counts as no more than one tie.}

Thirdly, we use the approach of \citet{mogstad2020, mogstad2022} to construct marginal confidence sets for the ranks. Unlike the methods proposed by \citet{xie2009} and \citet{hall2009},\footnote{\citet{hall2009} propose using an "$m$ out of $n$" bootstrap where the resample size, $m$, is smaller than the original sample size of the data, $n$. In our case, this would lead to more journals in the resampled data not citing any other journals in the sample, and, so, we did not pursue this approach.} that of \citet{mogstad2020} does not require the estimation or choice of tuning parameters.  First, we construct the following confidence set for each difference between pairs of recursive IFs:
\begin{equation}
\label{eq:mogstad}
   S_{i,j}\equiv [v_i-v_j\pm\hat \sigma_{i,j}s^{0.95}_i]
\end{equation}
where $\hat \sigma_{ij}$ is the estimated standard error of the difference between the recursive IFs, which we estimate as the standard deviation of the difference in the bootstrapped data. $s^{0.95}_i$ is the 0.95 quantile of
\begin{equation}
\label{eq:critval}
\max_{j: j\neq i} \frac{\vert v_{i,b}-v_{j,b}-(\bar v_i-\bar v_j)\vert}{\hat\sigma_{i,j}}
\end{equation}
where the $b$ indexes bootstrap iterations. We approximate the true difference between the recursive impact factors with the bootstrap mean difference $\bar v_i-\bar v_j$. In other words, for each bootstrap iteration we find the journal with the greatest absolute standardized difference in recursive IF to journal $i$ and select that value. Then we find the 95\% quantile $s^{0.95}_i$ of the values across all bootstrap iterations and use this in Equation (\ref{eq:mogstad}).

We use two calculations of $\hat \sigma_{i,j}$. First, we use ${\hat \sigma}_{i-j} = \sqrt{\hat \sigma_i^2 + \hat \sigma_j^2}$ as in \citet{mogstad2022}. This is a reasonable assumption for the application in \citet{mogstad2022} to the journal impact factors, which are independent by definition. However, the \emph{recursive} impact factor of journal $i$ depends on the recursive impact factors of all other journals, including journal $j$. We, therefore, also use $\hat \sigma_{i-j} = \sqrt{\hat \sigma_i^2 - 2\hat \sigma_{i,j} + \hat \sigma_j^2}$, where $\hat \sigma_i$, $\hat \sigma_j$, and $\hat \sigma_{i,j}$ are the standard deviations of $v_{i,b}$ and $v_{j,b}$ and the covariance of $v_{i,b}$ and $v_{j,b}$. 

Then we count the number of journals, $J^-_i$, whose differences with $i$ have a confidence set, $S_{i,j}$, that lies entirely below zero and the number of journals, $J^+_i$, whose differences with $i$ have a confidence set that lies entirely above zero. The set
\begin{equation}
\label{eq:mogset}
R_i^M\equiv \{J^-_i+1,...,J-J^+_i\}
\end{equation}
where $J$ is the total number of journals, covers the true rank of the journal with probability no less than $1-\alpha = 0.95$ \citep{mogstad2022}. That is, the Mogstad confidence interval is not narrower than the true confidence interval, but may be wider.

\section{Results}
\label{sc:results}
Table \ref{tab:results} presents the observed recursive IF, the standard error, and the 95\% confidence intervals for all journals. For the top ranked journals these confidence intervals are quite symmetric and closely match a classical confidence interval constructed using plus and minus 1.96 times the standard errors. However, as we go down the ranks the confidence intervals become increasingly asymmetric. Importantly, by construction, none of our estimated confidence intervals include negative numbers.

Our confidence intervals are much wider than those reported by \citet{lyhagen2020}. This is not surprising, as \citet{lyhagen2020} ignore the within journal variance. For example, for the \textit{American Economic Review} the upper value of their 95\% confidence interval is 102.7\% of their lower value. By contrast, our upper value is 122.8\% of our lower value. The relative sizes of our confidence intervals are similar to those estimated by \citet{stern2013} for the simple IFs. For the \textit{American Economic Review} he estimated an upper value that is 118.8\% of the lower value.

\begin{landscape}
\begin{figure}
    \centering
    \includegraphics[width=\textwidth]{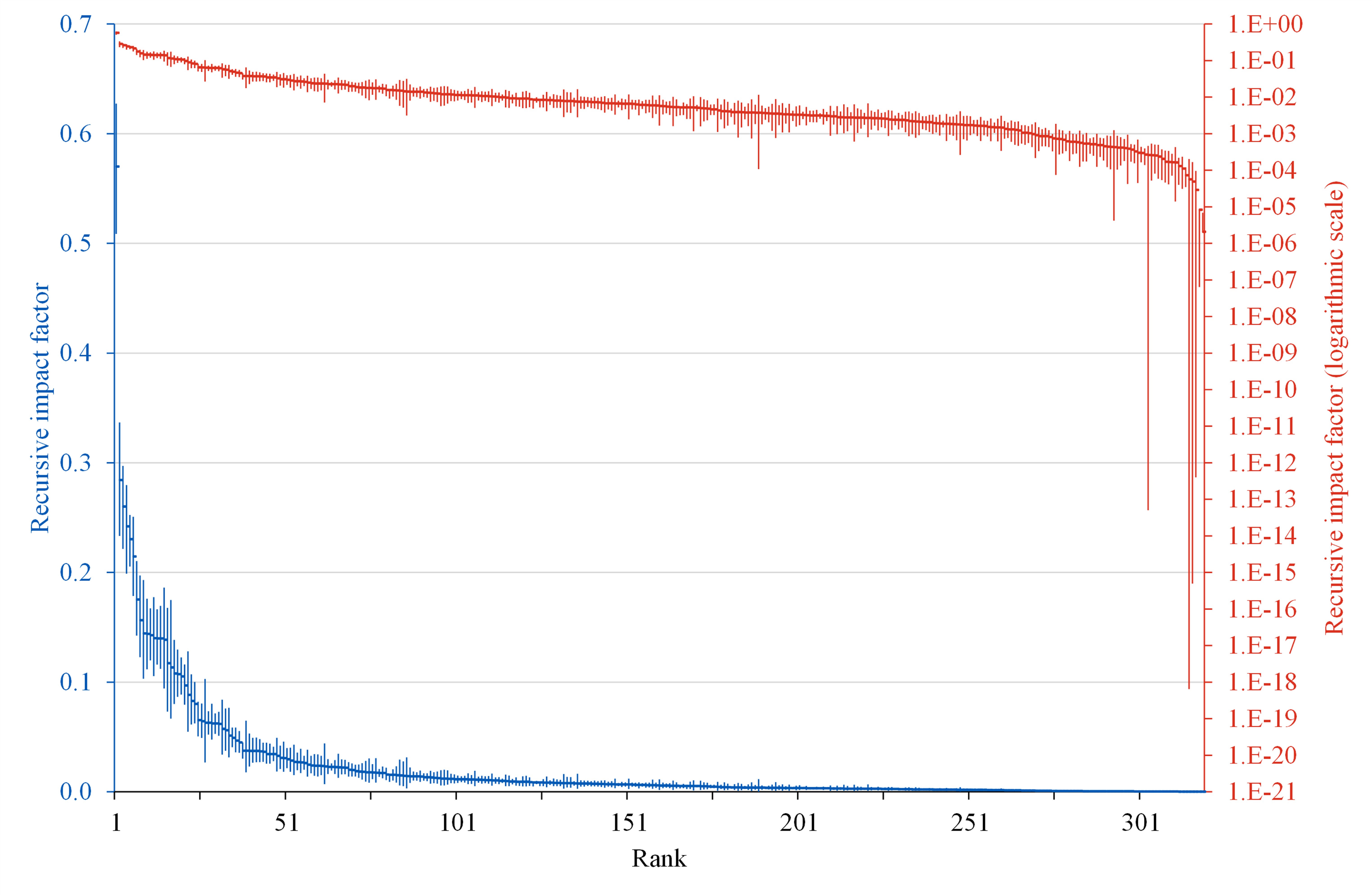}
    \caption{95\% confidence intervals of the recursive impact factor, arithmetic scale (left axis) and logarithmic scale (right axis)}
    \label{fig:rif}
\end{figure}
\end{landscape}

Figure \ref{fig:rif} presents the confidence intervals for all 319 journals. Clearly, the top ranked journal\textemdash the \textit{Quarterly Journal of Economics}\textemdash stands alone above all others. All other confidence intervals overlap. Figure \ref{fig:rif} presents the same data with a logarithmic scale so as to show detail for lower ranked journals.

Table \ref{tab:results} further shows the journal ranks based on the recursive impact factor and its 95\% confidence interval based on two alternative methods, the \citet{goldstein1996} method and the \citet{mogstad2022} method using $\hat \sigma_{i-j} = \sqrt{\hat \sigma_i^2 - 2\hat \sigma_{i,j} + \hat \sigma_j^2}$.. Table \ref{tab:rankings} also presents the confidence intervals using two variations of the \citet{xie2009} method and two further forms of the \citet{mogstad2022} method.

\begin{figure}
    \centering
    \includegraphics[width=\textwidth]{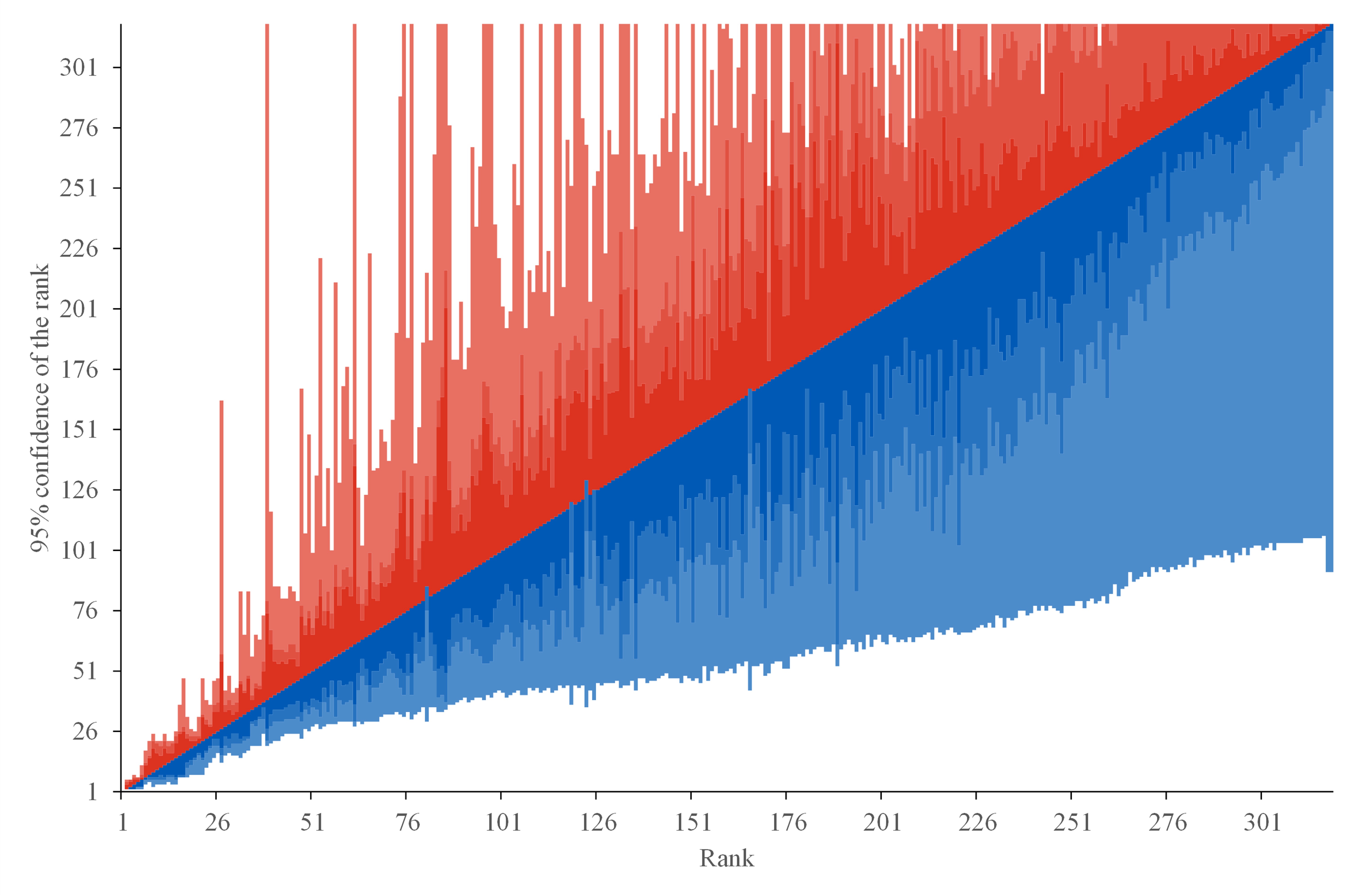}
    \caption{95\% confidence intervals of the rank based on the recursive impact factor. The inner intervals are based on Goldstein's bootstrap method, the middle intervals use Xie's correction to the bootstrap, and the outer intervals follow Mogstad's pairwise comparison.}
    \label{fig:rifrank}
\end{figure}

Figure \ref{fig:rifrank} shows these confidence intervals. The journals are ordered according to their empirical rank. The bootstrap confidence intervals are the narrowest. Mogstad's method yields the widest confidence intervals. Xie's approach lies in between.

The observed rank of five journals is outside the \citet{goldstein1996} confidence interval, specifically they rank higher empirically than the confidence interval suggests they should. These journals are all in the field of transportation and tend to cite each other. The most cited article published by these journals has 31 citations, 29 of which came from the 5 transportation journals. Dropping this article from some bootstrap samples probably generates this result.

The large standard errors notwithstanding, there is a clear ranking. Based on the narrower \citet{goldstein1996} bootstrap intervals, the four ``Top 5'' journals apart from the \textit{Quarterly Journal of Economics} are always ranked in the top 6, with the \textit{Journal of Finance} ranked between 3rd and 7th place. According to the widest Mogstad intervals, the ``Top 5'' are always in the top 8, with the \textit{AEJ Applied Economics} and the \textit{Journal of Labor Economics} perhaps in the top 5 too. \citet{hamermesh2018} found that citations to the \textit{Review of Economic Studies} have fallen over time compared to the other top-5 economics journals. This suggests that now maybe there is a top-4 group of economics journals. Our results do not support this. We find that the \textit{Review of Economic Studies} has a similar confidence interval for its rank as the \textit{American Economic Review}.

Mid-rank, the uncertainty is larger. It is largest for the \textit{Annals of Economics and Finance} (Goldstein intervals) and \textit{Economic Development and Cultural Change} (Mogstad intervals), which rank between 96th and 311st and between 20th and 319th, respectively. The uncertainty again narrows at the bottom. Using the Goldstein (Mogstad) intervals, the \textit{Korean Economic Review} (the \textit{Journal of Korea Trade}) never ranks higher than 316th (107th), out of 319.

The narrowing of the confidence intervals at the top and bottom of the rank order is a mechanical effect of the ranking process. At the top (bottom), a higher (lower) recursive impact factor does not lead to a higher (lower) rank. The two-sided uncertainty about the score translates into one-sided uncertainty about the rank.

Table \ref{tab:rankings} shows the confidence intervals for all six methods. The average width of the 95\% confidence interval for the \citet{goldstein1996} method is 53.4. Using the method of \citet[][Theorem 2.3]{xie2009} to resolve near-ties leads to considerably wider confidence intervals, with an average of 97.6 ranks. The pairwise method of \citet{mogstad2022} leads to even wider confidence intervals, on average 193.3.\footnote{Note that for 8 journals, all in the top 15, Xie's lower bound falls below Mogstad's.} If we correct their method for the correlation between journals (as shown in Table \ref{tab:results}), the average confidence interval widens slightly to 193.7.

The Xie confidence intervals are almost as wide, 182.8 on average, as the Mogstad confidence intervals if we use Mogstad's bandwidth (the standard deviation of the difference in scores) with Xie's method. However, using Xie's bandwidth (the interquartile range of scores times the square root of the standard deviation of the score of the journal of interest) with Mogstad's method widens Mogstad's confidence intervals to an average of 251.1. Both \citet{xie2009} and \citet{mogstad2022} have chosen their bandwidth to reduce type II errors.

\section{Conclusions}
\label{sc:conclusion}
Journal rankings are often interpreted as if the underlying performance indicators are measured with great precision. This is not the case. We present here confidence intervals for the recursive impact factor and the associated rank for 319 economics journals. The recursive impact factor improves on the journal impact factor as it puts more weight on citations in more prestigious journals. This recursivity makes it harder to compute confidence intervals. The methods proposed here improve on the earlier proposal by \citet{lyhagen2020} in that we account for within journal variation in citations and we use estimates of confidence intervals for ranks that allow for the effect of ties and close ties. The resulting confidence intervals are wide: Using the narrowest confidence intervals produced by the \citet{goldstein1996} method, mid-ranking journals could just as well have ranked 25 places higher or lower, while using the much broader confidence intervals produced by the \citet{mogstad2022} method, such journals could rank more than 90 places higher or lower. However, confidence intervals on top and bottom journals are much narrower. The \textit{Quarterly Journal of Economics} is indisputably the most influential journal, followed by the rest of the traditional Top 5 and the \textit{Journal of Finance}.

We also show that there are considerable differences in the results of the various methods for constructing confidence intervals for ranks. The simple bootstrap proposed by \citet{goldstein1996} may appeal because of its simplicity and small-sample properties, but ties and near-ties between journal imply that the confidence intervals are too narrow. \citet{mogstad2022} propose a method to correct for this that does not require additional parametric assumptions but their confidence intervals may be too wide by construction. The correction proposed by \citet{xie2009} is a consistent estimator of the confidence intervals, but relies on assumptions about kernel functions and bandwidths.

The journal ranking presented here should be interpreted with caution. Not just because any ranking should, and not just because the confidence intervals are wide, but also because we only consider citations from and to journals \textit{within} the discipline of economics. Economics journals that are frequently cited in journals of cognate disciplines are, therefore, discounted. This particularly affects economics journals on the boundaries with law, psychology, policy, or the environment.

\begin{landscape}
\footnotesize
\begin{longtable}{|r|l|c|r|r|r|r|r|r|r|r|}
\caption{Key statistics for the recursive impact factor of economics journals.} \label{tab:results} \\

\hline \textbf{Rank} & \textbf{Journal} & \textbf{ISSN} & \multicolumn{4}{|c|}{\textbf{Recursive impact factor}} & \multicolumn{2}{|c|}{\textbf{Goldstein}} & \multicolumn{2}{|c|}{\textbf{Mogstad}}\\
 &  &  & \textbf{Mean} &
\textbf{S.E.} & \textbf{lower} & \textbf{upper} & \textbf{lower} & \textbf{upper} & \textbf{lower} & \textbf{upper}\\\hline 
\endfirsthead

\multicolumn{11}{c}%
{{\bfseries \tablename\ \thetable{} -- continued from previous page}} \\
\hline \textbf{Rank} & \textbf{Journal} & \textbf{ISSN} & \multicolumn{4}{|c|}{\textbf{Recursive impact factor}} & \multicolumn{2}{|c|}{\textbf{Goldstein}} & \multicolumn{2}{|c|}{\textbf{Mogstad}}\\
 &  &  & \textbf{Mean} &
\textbf{S.E.} & \textbf{lower} & \textbf{upper} & \textbf{lower} & \textbf{upper} & \textbf{lower} & \textbf{upper}\\ \hline 
\endhead

\hline \multicolumn{11}{|r|}{{Continued on next page}} \\ \hline
\endfoot

\hline \hline
\endlastfoot

1	&	QUARTERLY JOURNAL OF ECONOMICS 	&	 0033-5533 	&	0.570030	&	0.031130	&	0.508630	&	0.627330	&	1	&	1	&	1	&	1	\\
2	&	ECONOMETRICA 	&	 0012-9682 	&	0.284210	&	0.026560	&	0.233370	&	0.336640	&	2	&	4	&	2	&	6	\\
3	&	JOURNAL OF POLITICAL ECONOMY 	&	 0022-3808 	&	0.259970	&	0.019550	&	0.221500	&	0.297050	&	2	&	5	&	2	&	6	\\
4	&	REVIEW OF ECONOMIC STUDIES 	&	 0034-6527 	&	0.241850	&	0.020320	&	0.198920	&	0.279650	&	2	&	6	&	2	&	8	\\
5	&	AMERICAN ECONOMIC REVIEW 	&	 0002-8282 	&	0.230140	&	0.012040	&	0.205480	&	0.252330	&	3	&	6	&	2	&	7	\\
6	&	JOURNAL OF FINANCE 	&	 0022-1082 	&	0.214530	&	0.018870	&	0.178490	&	0.250840	&	3	&	7	&	2	&	12	\\
7	&	\thead{AMERICAN ECONOMIC JOURNAL\\APPLIED ECONOMICS} 	&	 1945-7782 	&	0.175250	&	0.017230	&	0.142440	&	0.210250	&	6	&	12	&	4	&	18	\\
8	&	JOURNAL OF LABOR ECONOMICS 	&	 0734-306X 	&	0.156380	&	0.018680	&	0.122640	&	0.197190	&	7	&	15	&	5	&	22	\\
\end{longtable}

\end{landscape}

\bibliography{Recursive}

\newpage \appendix
\setcounter{table}{0}
\renewcommand{\thetable}{A\arabic{table}}

\begin{table}[]
    \centering
    \begin{tabular}{c l c}
         Symbol & Description & Dimension \\ \hline
         $A$ & Total number of papers published by journal & $J \times J$ (diagonal)\\
         $B$ & Number of bootstraps & 1 \\
         $b$ & Index for bootstrap & 1 \\
         $C$ & Journal by journal citation matrix & $J \times J$ \\
         $D$ & Total number of citations received by journal & $J \times J$ (diagonal) \\
         $F$ & Cumulative distribution function & 1 \\
         $f$ & Probability density function & 1 \\
         $i$ & Index for cited journal of interest & 1 \\
         $J$ & Number of journals & 1 \\ 
         $j$ & Index for citing journal & 1 \\
         $K$ & Number of papers & 1 \\
         $k$ & Index for paper & 1 \\
         $P$ & Paper by journal citation matrix & $K \times J$ \\
         $Q$ & Journal by paper concordance tables & $J \times K$\\
         $R$ & Journal rank & $J \times 1$ \\
         $T$ & Number of ties & $J \times 1$ \\
         $v$ & Recursive impact factor & $J \times 1$ \\
         $\tau$ & Bandwidth & $J \times 1$ or $J \times J$ \\
         $\hat{\sigma}$ & Estimated standard deviation & $J \times 1$ or $J \times J$\\
         $s^\alpha$ & Critical value for confidence level $\alpha$ & $J \times 1$
         \\ \hline
    \end{tabular}
    \caption{Notation}
    \label{tab:notation}
\end{table}

\begin{landscape}
\footnotesize
\begin{longtable}{|r l|r r|r r|r r|r r|r r|r r|}
\caption{Alternative confidence intervals for the rankings on the recursive impact factor of economics journals.} \label{tab:rankings} \\

\hline \textbf{} & \textbf{Journal} & \multicolumn{2}{|c|}{\textbf{Goldstein}} & \multicolumn{2}{|c|}{\textbf{Xie}} & \multicolumn{2}{|c|}{\textbf{Xie/M}} & \multicolumn{2}{|c|}{\textbf{Mogstad}} & \multicolumn{2}{|c|}{\textbf{Mogstad+}} & \multicolumn{2}{|c|}{\textbf{Mogstad/X}}\\
 &  & \textbf{low} & \textbf{high}  & \textbf{low} & \textbf{high} & \textbf{low} & \textbf{high}  & \textbf{low} & \textbf{high} & \textbf{low} & \textbf{high} & \textbf{low} & \textbf{high}\\\hline 
\endfirsthead

\multicolumn{14}{c}%
{{\bfseries \tablename\ \thetable{} -- continued from previous page}} \\
\hline \textbf{} & \textbf{Journal} & \multicolumn{2}{|c|}{\textbf{Goldstein}} & \multicolumn{2}{|c|}{\textbf{Xie}} & \multicolumn{2}{|c|}{\textbf{Xie/M}} & \multicolumn{2}{|c|}{\textbf{Mogstad}} & \multicolumn{2}{|c|}{\textbf{Mogstad+}} & \multicolumn{2}{|c|}{\textbf{Mogstad/X}}\\
 &  & \textbf{low} & \textbf{high} & \textbf{low} & \textbf{high} & \textbf{low} & \textbf{high} & \textbf{low} & \textbf{high} & \textbf{low} & \textbf{high} & \textbf{low} & \textbf{high}\\ \hline 
\endhead

\hline \multicolumn{14}{|r|}{{Continued on next page}} \\ \hline
\endfoot

\hline \hline
\endlastfoot
1	&	QUARTERLY JOURNAL OF ECONOMICS 	&	1	&	1	&	1	&	1	&	1	&	2	&	1	&	1	&	1	&	1	&	1	&	1	\\
2	&	ECONOMETRICA 	&	2	&	4	&	2	&	5	&	1	&	7	&	2	&	6	&	2	&	6	&	2	&	6	\\
3	&	JOURNAL OF POLITICAL ECONOMY 	&	2	&	5	&	2	&	6	&	1	&	8	&	2	&	6	&	2	&	6	&	2	&	6	\\
4	&	REVIEW OF ECONOMIC STUDIES 	&	2	&	6	&	2	&	7	&	1	&	9	&	2	&	8	&	2	&	8	&	2	&	8	\\
5	&	AMERICAN ECONOMIC REVIEW 	&	3	&	6	&	3	&	7	&	1	&	9	&	2	&	7	&	2	&	7	&	2	&	7	\\
6	&	JOURNAL OF FINANCE 	&	3	&	7	&	3	&	8	&	1	&	12	&	2	&	12	&	2	&	12	&	2	&	15	\\
7	&	\thead{AMERICAN ECONOMIC JOURNAL\\APPLIED ECONOMICS} 	&	6	&	12	&	6	&	13	&	1	&	19	&	4	&	18	&	4	&	18	&	4	&	20	\\
8	&	JOURNAL OF LABOR ECONOMICS 	&	7	&	15	&	6	&	16	&	1	&	23	&	5	&	22	&	5	&	22	&	5	&	22	\\
\end{longtable}

\footnotesize \textit{Notes:} Journals are ordered by their recursive impact factor. The columns ``Goldstein'' show the bootstrap confidence intervals of the ranks. The columns ``Xie'' use the correction of the bootstrap confidence intervals proposed by \citet{xie2009}. The columns ``Xie/M'' use the same correction, but with Mogstad's bandwidth instead. The columns ``Mogstad'' use the method proposed by \citet{mogstad2020} to compute confidence intervals. The columns ``Mogstad+'' use the same method corrected for correlations between recursive impact factors. The columns ``Mogstad/X'' use the same method again but using Xie's bandwidth instead. Columns ``Goldstein'' and ``Mogstad+'' are also shown in Table \ref{tab:results}.
\end{landscape}
\end{document}